\documentclass[final,3p,times]{elsarticle}

\usepackage{amssymb}
\usepackage{amsmath}
\journal{Journal of Subatomic Particles and Cosmology}

\begin{document}

\begin{frontmatter}

%% Title, authors and addresses

%% use the tnoteref command within \title for footnotes;
%% use the tnotetext command for theassociated footnote;
%% use the fnref command within \author or \affiliation for footnotes;
%% use the fntext command for theassociated footnote;
%% use the corref command within \author for corresponding author footnotes;
%% use the cortext command for theassociated footnote;
%% use the ead command for the email address,
%% and the form \ead[url] for the home page:
%% \title{Title\tnoteref{label1}}
%% \tnotetext[label1]{}
%% \author{Name\corref{cor1}\fnref{label2}}
%% \ead{email address}
%% \ead[url]{home page}
%% \fntext[label2]{}
%% \cortext[cor1]{}
%% \affiliation{organization={},
%%             addressline={},
%%             city={},
%%             postcode={},
%%             state={},
%%             country={}}
%% \fntext[label3]{}

\title{Demonstrating CBM Capabilities through $\Lambda$ Reconstruction with mCBM}

%% use optional labels to link authors explicitly to addresses:
\author[aaa]{Yue Hang Leung for the CBM collaboration}
\affiliation[aaa]{organization={Physikalisches Institut, Universität Heidelberg},
%             addressline={111 Abc St.},
             city={Heidelberg},
%             postcode={12345},
%             state={ZZ},
             country={Germany}}

% \affiliation[bbb]{organization={YYY Laboratory},
%             addressline={999 Def Rd.},
%             city={NNN},
%             postcode={56789},
%             state={YY},
%             country={USA}}

%% Abstract
\begin{abstract}
%% Text of abstract
The Compressed Baryonic Matter (CBM) experiment at the upcoming Facility for Antiproton and Ion Research (FAIR) is a high-rate fixed-target experiment designed to investigate nuclear matter at extreme baryon densities in relativistic nucleus-nucleus collisions. To enable high-statistics measurements of rare probes, CBM is designed to operate at event rates up to 10 MHz. This necessitates the development of fast and radiation-tolerant detectors, self-triggered front-end electronics, a free-streaming data acquisition architecture, and real-time event reconstruction capabilities. Prototype versions and pre-series productions of the CBM detector systems have been deployed in the mini-CBM demonstrator setup mCBM — an experimental precursor comprising sub-components of all major CBM systems, installed at the SIS18 facility of GSI/FAIR within the FAIR Phase-0 program.

In 2024, Ni$+$Ni collisions at a kinetic beam energy of 1.93 AGeV and an average interaction rate of about 500 kHz were successfully recorded. This dataset enables a detailed evaluation of the operational performance of the detector systems as well as the complete CBM data chain, while the reconstruction of $\Lambda$ baryons, which constitute a relatively rare signal within the limited geometrical acceptance of mCBM, serves as a natural benchmark. This paper presents the first results on $\Lambda$ baryon reconstruction with the mCBM experiment, demonstrating the functionality of the detector technologies and the data chain for the upcoming full-scale CBM experiment.
\end{abstract}

%%Graphical abstract
%\begin{graphicalabstract}
%\includegraphics{grabs}
%\end{graphicalabstract}

%%Research highlights
%\begin{highlights}
%\item Research highlight 1
%\item Research highlight 2
%\end{highlights}

%% Keywords
\begin{keyword}
%% keywords here, in the form: keyword \sep keyword

%% PACS codes here, in the form: \PACS code \sep code

%% MSC codes here, in the form: \MSC code \sep code
%% or \MSC[2008] code \sep code (2000 is the default)

\end{keyword}

\end{frontmatter}

%% Add \usepackage{lineno} before \begin{document} and uncomment 
%% following line to enable line numbers
%% \linenumbers

%% main text
%%

%% Use \section commands to start a section
\section{Introduction}
\label{sec1}
%% Labels are used to cross-reference an item using \ref command.

The Compressed Baryonic Matter (CBM) experiment~\cite{cbm} at the Facility for Antiproton and Ion Research (FAIR)~\cite{fair}
is designed to investigate strongly interacting matter at high net-baryon densities in relativistic heavy-ion
collisions~\cite{CbmPhyProg2017}. To measure rare probes with high precision, CBM will operate at interaction rates
of up to 10 MHz using a triggerless, free-streaming data acquisition system, in which detector data are
continuously reconstructed online.

To validate this experimental concept, the mini-CBM (mCBM) demonstrator~\cite{mCBM-proposal2017,mCBM-proposal2020,mCBM-proposal2022,CBM:2026kqm}
operates at the SIS18 accelerator within the FAIR Phase-0 program, employing prototype and pre-series CBM detector
systems under realistic high-rate conditions. The primary aim of these proceedings is to demonstrate the operation of the mCBM experiment as a full-system precursor of the CBM experiment, encompassing detector integration, free-streaming data transport, and particle reconstruction
using the CBM data-acquisition and reconstruction framework. To this end, the reconstruction of $\Lambda$ baryons is employed as a benchmark analysis, as it represents a relatively rare signal within the limited geometrical acceptance of mCBM, with an integral acceptance of approximately $6.5\times10^{-3}$, and has a characteristic weak-decay topology requiring efficient track reconstruction and effective background suppression in a high-rate, triggerless environment. 

\section{The mCBM experimental setup and data transport system}
\label{sec:mcbm_setup}

\begin{figure}[h!] 
\centering
\includegraphics[width=0.75\linewidth]{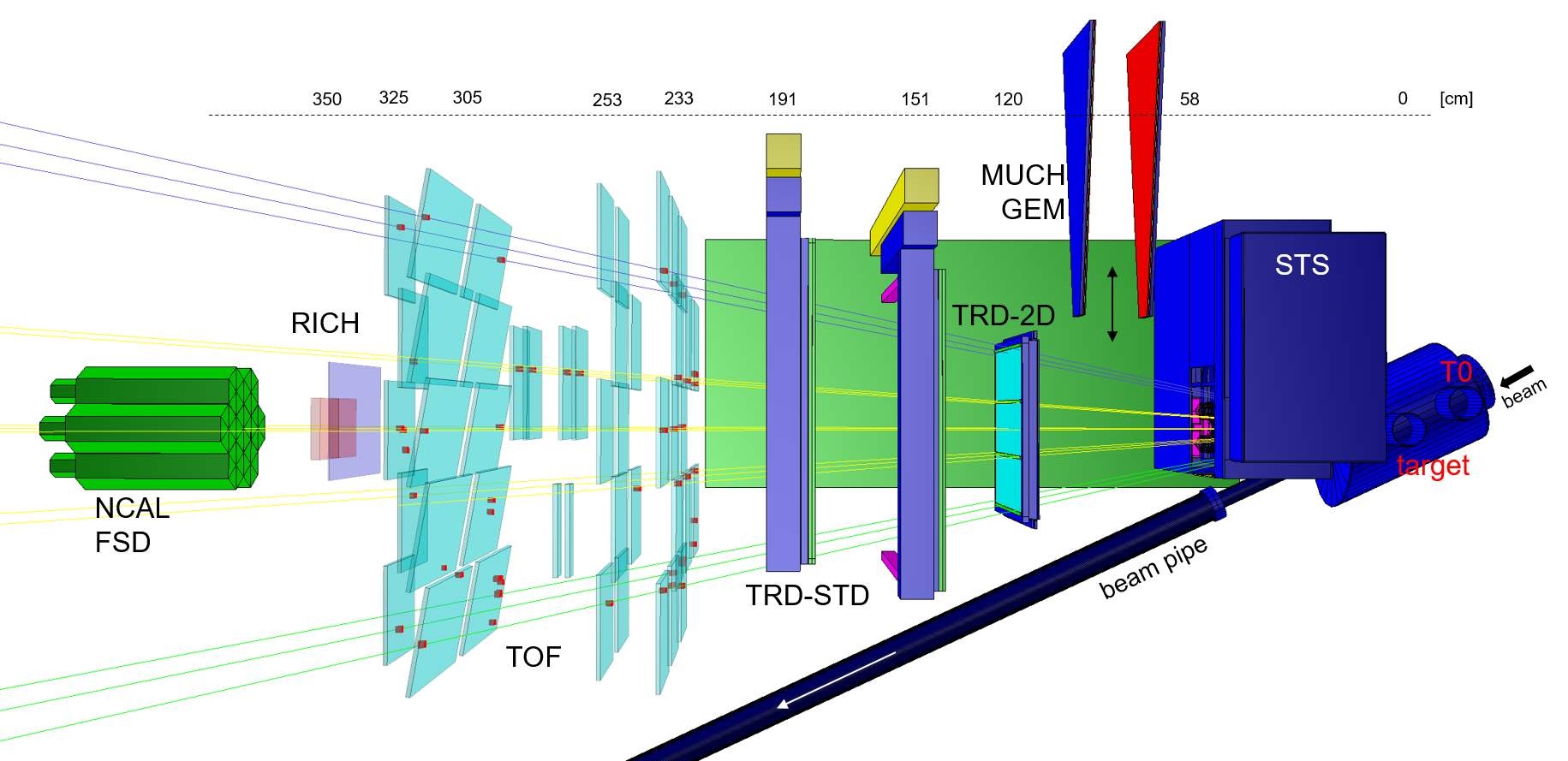}
\caption{Top view ROOT geometry of the 2024 mCBM setup. }
\label{fig:mcbm_setup_2024}
\end{figure}

Figure~\ref{fig:mcbm_setup_2024} shows the mCBM setup used during the 2024 SIS18 beam campaign. The experiment employs prototype and pre-series CBM detector systems arranged downstream of a fixed target at a polar angle of about $25^\circ$. A detailed description of the detector systems can be found in Refs.~\cite{mCBM-proposal2022,CBM:2025voh}. The present analysis primarily uses the Beam Monitoring system (BMON), the Silicon Tracking System (STS), and the Time-of-Flight (TOF) detector. The BMON provides the collision start time, the three-station STS reconstructs charged-particle tracks and decay vertices, and the TOF detector provides timing measurement. The remaining detector systems (MUCH, TRD, RICH, and FSD/NCAL) participated in the data taking but are not used in the present analysis.

%\subsection{The CBM data transport system under test at mCBM}
%\label{daq}

%\begin{figure}[pos=b]
%\centering
%\includegraphics[width=0.95\linewidth]{1-IntroSetup/figs/1-3-2024-CBM-data-chain-noPhoto.png}
%\caption{Schematic of the triggerless CBM data transport chain as implemented in mCBM.}
%\label{fig:DataChain}
%\end{figure}

The CBM data acquisition system, used in the mCBM experiment, employs a triggerless, free-streaming architecture. Detector front-end electronics continuously digitize signals above threshold, assign precise timestamps synchronized by the Timing and Fast Control (TFC) system, and transmit the data via GBT links to Common Readout Interface (CRI) boards. The First-Level Event Selector (FLES) assembles detector data into timeslices. The mCBM experiment serves as a demonstrator of this architecture under realistic high-rate conditions. During the 2024 campaign, the FLES continuously recorded detector data at rates of up to about 1 GB/s. 

\section{Data analysis}
\label{sec:DataAna}

The analysis was performed within the CbmRoot framework~\cite{Al-Turany:2012zfk}. The dataset consists of Ni+Ni collisions at a beam kinetic energy of 1.93\,AGeV ($\sqrt{s_{\rm NN}}=2.67$\,GeV), recorded during the 2024 mCBM campaign. Event candidates are defined by a software event-building algorithm. A coincidence of four TOF digis from two MRPCs together with a matching BMON T0 hit defines an event. Additional TOF, BMON, and STS digis are then assigned to the event using fixed time windows. Approximately $2.9\times10^{9}$ events were identified and analyzed. 

\begin{figure}[h!] 
\centering
\includegraphics[width=.59\linewidth]{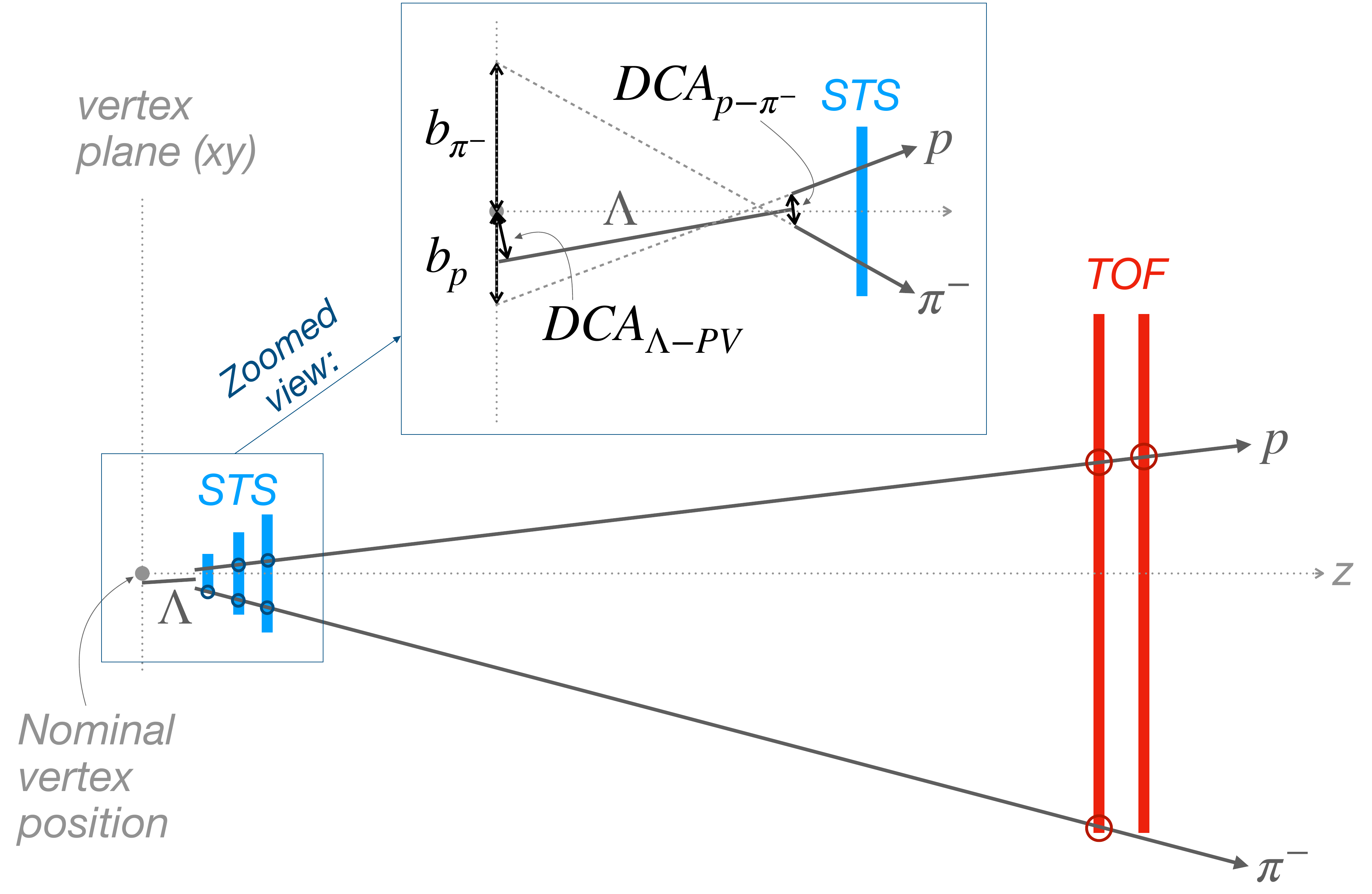} 
\includegraphics[width=.395\linewidth]{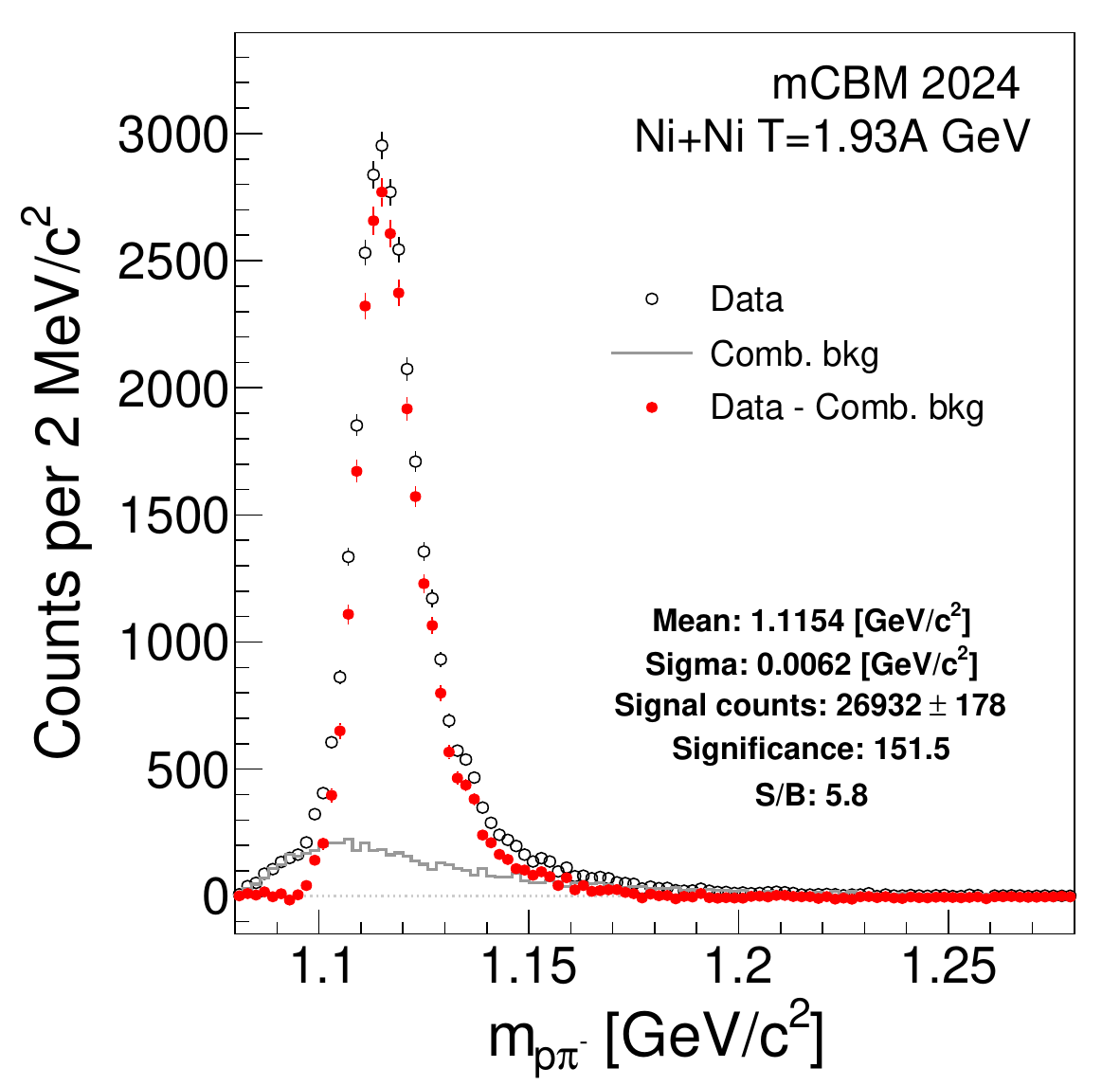} 
\caption{(left) Schematic of $\Lambda$ decay reconstruction in the mCBM setup. The proton track is reconstructed using two STS hits and two TOF hits, whereas the pion track is reconstructed using three STS hits and one TOF hit. (right) Invariant mass distribution of $p$-$\pi^{-}$ pairs. Open circles represent the data, the grey histogram represents the combinatorial background, and the red dots represent the background-subtracted data.}
\label{fig:lambda_inv_mass}
\end{figure}

As shown in Fig.~\ref{fig:lambda_inv_mass}, $\Lambda$ candidates are reconstructed through the decay $\Lambda\rightarrow p+\pi^{-}$. Before reconstructed tracks can be used for $\Lambda$ reconstruction, the detector timing and geometry must be calibrated. TOF calibration corrects for signal propagation, time walk, and channel-dependent timing offsets, the latter being determined from the speed-of-light edge of the TOF distributions with respect to the BMON start time. The resulting average MRPC time resolution is about 90 ps. Detector positions are refined using a data-driven iterative alignment procedure that minimizes the residuals between reconstructed tracks and detector hits.

After TOF calibration and detector alignment, charged-particle tracks are reconstructed in the field-free geometry of mCBM using the Cellular Automaton algorithm. Since the tracks are reconstructed as straight lines, their momenta cannot be determined from track curvature. Instead, the particle velocity is calculated from the measured time of flight and reconstructed flight path, and the momentum is obtained under either the proton or pion mass hypothesis. A track is assigned the proton (pion) mass hypothesis if it satisfies the corresponding velocity and transverse impact parameter requirements. 

The proton and pion candidates are then combined to reconstruct $\Lambda$ baryons, using the KFParticle package~\cite{Zyzak2016}. Topological selections on the transverse impact parameter of the daughter particles, the DCA between daughters, and the DCA between the mother and the primary vertex, are applied to suppress combinatorial background. The combinatorial background is estimated using an event-mixing technique. After background subtraction, the $\Lambda$ signal is extracted by counting candidates in the invariant-mass window around the nominal $\Lambda$ mass. Figure~\ref{fig:lambda_inv_mass} shows the resulting invariant-mass distribution, from which $(2.69\pm0.02)\times10^{4}$ $\Lambda$ candidates are reconstructed.

\section{Simulation and validation}
\label{sec:sim}

In order to validate that the observed signal corresponds to real $\Lambda$ decays, as well as to determine the acceptance and efficiency necessary for lifetime and yield measurements, simulations of $\Lambda$ decays are carried out within the CbmRoot framework. Generated particles were transported through the full mCBM geometry using GEANT3~\cite{Brun:1994aa}, followed by detector digitization, hit reconstruction, track reconstruction, and $\Lambda$ reconstruction using the same software chain as for experimental data. The simulation includes the measured beam profile, detector time resolution, dead channels, detector inefficiencies, and residual detector misalignment. The $\Lambda$ source was parameterized using published FOPI measurements~\cite{FOPI:2007usx}, while PHQMD simulations were used to adjust the Boltzmann and Gaussian source parameters to account for the centrality selection of the mCBM dataset. PHQMD has been shown to reproduce charged-particle multiplicities sufficiently well at comparable collision energies~\cite{Zhou:2025zgn}. Acceptance and reconstruction efficiencies were determined using simulated $\Lambda$ decays embedded into PHQMD events. 

It is found that the simulation reproduces the transverse momentum distribution, the transverse impact parameter of the daughters, and the various DCA quantities in the background-subtracted data well. These comparisons demonstrate that the simulation provides a realistic description of the detector response and $\Lambda$ reconstruction performance, supporting its use for efficiency corrections and the extraction of physics observables.

\iffalse
\begin{figure}[t]
\centering
\includegraphics[width=.32\linewidth]{4-3-hypt_2d_mc_py.pdf}
\includegraphics[width=.32\linewidth]{4-6-dca_pion_mc.pdf}
\includegraphics[width=.32\linewidth]{4-8-dca_lambda_mc.pdf}
\caption{Comparison between background-subtracted data (red) and simulation (blue) for representative observables: (left) transverse momentum, (center) pion transverse impact parameter $b_{\pi^-}$, and (right) distance of closest approach of the reconstructed $\Lambda$ candidate to the primary vertex, $DCA_{\Lambda-\mathrm{PV}}$.}
\label{fig:mc_validation}
\end{figure}
\fi

\section{Results}
\label{sec:results}

\subsection{$\Lambda$ lifetime}

The proper decay length is calculated as $L/(\beta\gamma)=Lm/p$, where $L$ is the reconstructed decay length, $p$ the $\Lambda$ momentum obtained from KFParticle, and $m$ the PDG $\Lambda$ mass~\cite{ParticleDataGroup:2024cfk}. The raw signal yield is extracted in bins of proper decay length. After correcting for the acceptance and reconstruction efficiency determined from simulation, the proper decay length distribution is fitted with an exponential function to extract the lifetime. 

\begin{figure}[t]
\centering
\begin{minipage}[c]{0.40\linewidth}
    \centering
    \includegraphics[width=\linewidth]{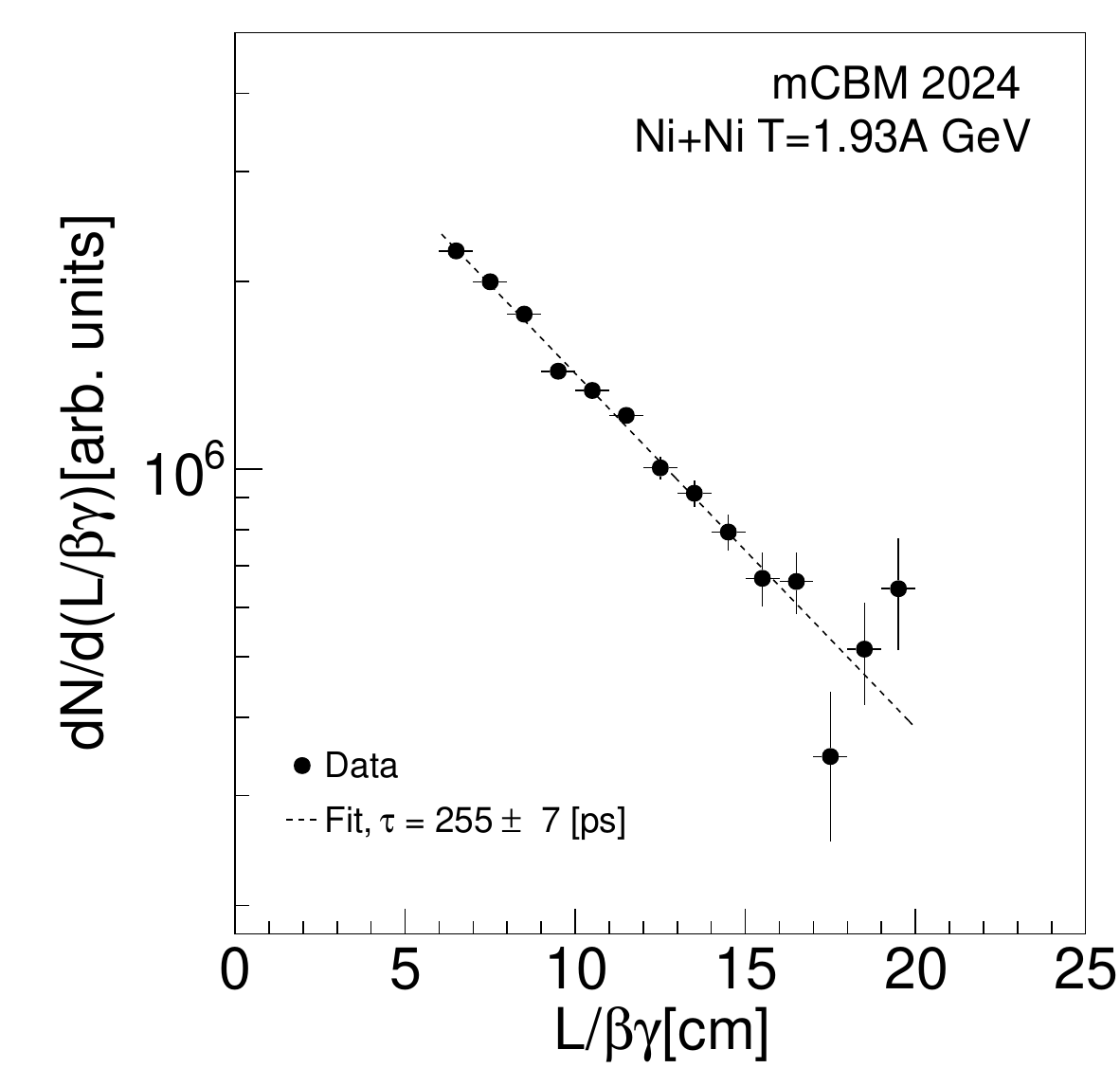}
\end{minipage}
\hfill
\begin{minipage}[c]{0.55\linewidth}
    \caption{Acceptance and efficiency corrected proper decay length distribution of reconstructed $\Lambda$ baryons. The dashed line shows the exponential fit.}
    \label{fig:lambda_lifetime}
\end{minipage}
\end{figure}

\iffalse
\begin{figure}[t]
\centering
\includegraphics[width=.4\linewidth]{4-11-lambda_lifetime_0_avg.pdf}
\caption{Acceptance and efficiency corrected proper decay length distribution of reconstructed $\Lambda$ baryons. The dashed line shows the exponential fit.}
\label{fig:lambda_lifetime}
\end{figure}
\fi

Figure~\ref{fig:lambda_lifetime} shows the corrected proper decay length distribution together with the exponential fit. The systematic uncertainty was evaluated by varying the signal extraction procedure, detector response, topological selections, and the simulated $\Lambda$ kinematic distributions. The extracted lifetime is $\tau=(255\pm7~\mathrm{(stat.)}\pm28~\mathrm{(syst.)})~\mathrm{ps}$, consistent with the PDG average of $(263\pm2)$\,ps. 

\subsection{$\Lambda$ multiplicity}

The acceptance and reconstruction efficiency determined from the simulation amounts to $\epsilon_{\rm acc}=(6.5\pm0.7)\times10^{-3}$ and $\epsilon_{\rm reco}=(4.6\pm0.9)\times10^{-2}$, respectively.
The measured $\Lambda$ multiplicity in the triggered event sample, which is close to minimum-bias, is extrapolated to $4\pi$ phase space and amounts to $M_{\Lambda,\rm trig}=0.031\pm0.0002~\mathrm{(stat.)}$. As a cross-check, the triggered-event multiplicity is extrapolated to central collisions corresponding to a geometrical cross section of $350\,\mathrm{mb}$ using a PHQMD-derived scaling factor of $2.94\pm0.30$, enabling a direct comparison with the published FOPI result~\cite{FOPI:2007usx}. The dominant systematic uncertainties arise from the acceptance and reconstruction efficiency and the extrapolation to the unmeasured phase space. The resulting $4\pi$-phase-space $\Lambda$ multiplicity is $M_{\Lambda,\rm cen}=0.091\pm0.0006~\mathrm{(stat.)}\pm0.025~\mathrm{(syst.)}$,
consistent within $1.3\sigma$ with the published FOPI measurement of $M_{\Lambda,\rm cen}=0.137\pm0.005~\mathrm{(stat.)}^{+0.009}_{-0.025}~\mathrm{(syst.)}$.

\section{Conclusions and Outlook}
\label{sec:summary}

The mCBM experiment serves as a demonstrator of the detector technologies, triggerless data acquisition, and reconstruction framework being developed for the CBM experiment at FAIR. In this work, $\Lambda\rightarrow p+\pi^{-}$ decays were successfully reconstructed in Ni+Ni collisions at a beam kinetic energy of 1.93\,AGeV using the CBM data-acquisition and reconstruction framework. A statistically significant signal was successfully extracted, with the topological and kinematic distributions agreeing with GEANT simulations and the measured $\Lambda$ lifetime and multiplicity being consistent with published values. These results demonstrate that non-trivial and relatively rare signals within the detector acceptance can be reliably reconstructed from
continuous, triggerless data using the CBM data acquisition and reconstruction framework, providing an important validation of the experimental concept of CBM. The experience gained with mCBM, together with the large data samples collected during the 2022--2025 campaigns, will support the commissioning and optimization of reconstruction algorithms in preparation for the start of CBM data taking at FAIR.
%% If you have bib database file and want bibtex to generate the
%% bibitems, please use
%%
\bibliographystyle{elsarticle-num}
\bibliography{sqm2026_template}

%% else use the following coding to input the bibitems directly in the
%% TeX file.

%% Refer following link for more details about bibliography and citations.
%% https://en.wikibooks.org/wiki/LaTeX/Bibliography_Management

\end{document}